\documentclass[a4paper,12pt]{article}
\usepackage{graphicx,rotating,amsmath,multirow,xcolor,colortbl,xcolor}
\usepackage{hyperref,pdflscape,slashed,charter}
\usepackage{makecell}
\pdfoutput=1
\hypersetup{colorlinks,bookmarksopen,bookmarksnumbered,citecolor=verdes,
linkcolor=blus,pdfstartview=FitH,urlcolor=rossos}

\allowdisplaybreaks

\def\hhref#1{\href{http://arxiv.org/abs/#1}{arXiv:#1}} 
\def\arXiv#1{\href{http://arxiv.org/abs/#1}{arXiv:#1}} 

\usepackage{multicol}
\usepackage{color}
\definecolor{rosso}{cmyk}{0,1,1,0.4}
\definecolor{rossos}{cmyk}{0,1,1,0.55}
\definecolor{rossoc}{cmyk}{0,1,1,0.2}
\definecolor{blu}{cmyk}{1,1,0,0.3}
\definecolor{blus}{cmyk}{1,1,0,0.6}
\definecolor{bluc}{cmyk}{1,1,0,0.1}
\definecolor{verde}{cmyk}{0.92,0,0.59,0.25}
\definecolor{verdec}{cmyk}{0.92,0,0.59,0.15}
\definecolor{verdes}{cmyk}{0.92,0,0.59,0.4}
\definecolor{verdess}{cmyk}{0.92,0,0.59,0.8}

\def\eq#1{eq.~(\ref{#1})}

\definecolor{Gray}{gray}{0.95}

\newcommand{\GeV}{\,{\rm GeV}}
\newcommand{\TeV}{\,{\rm TeV}}

\newcommand{\Tr}{\,{\rm Tr}}

\def\circa#1{\,\raise.3ex\hbox{$#1$\kern-.75em\lower1ex\hbox{$\sim$}}\,}

\newcommand{\beq}{\begin{equation}}
\newcommand{\eeq}{\end{equation}}

\newcommand{\bea}{\begin{eqnarray}}
\newcommand{\eea}{\end{eqnarray}}
\newcommand{\be}{\begin{equation}}
\newcommand{\ee}{\end{equation}}
\font\tenrsfs=rsfs10 at 12pt
\font\sevenrsfs=rsfs7
\font\fiversfs=rsfs5
\newfam\rsfsfam
\textfont\rsfsfam=\tenrsfs
\scriptfont\rsfsfam=\sevenrsfs
\scriptscriptfont\rsfsfam=\fiversfs
\def\mathscr#1{{\fam\rsfsfam\relax#1}}
\def\Lag{\mathscr{L}}

\newcommand{\ba}{\begin{array}}
\newcommand{\ea}{\end{array}}

\def\Re{{\rm Re}\,}

\renewcommand{\theequation}{\thesection.\arabic{equation}}

\newcommand{\mysection}[1]{\section{#1}\setcounter{equation}{0}}

\def\circa#1{\,\raise.3ex\hbox{$#1$\kern-.75em\lower1ex\hbox{$\sim$}}\,}
\makeatletter

\def\art{\@ifnextchar[{\eart}{\oart}}
\def\eart[#1]#2#3#4#5#6{{\rm #2}, {\em #3 \bf #4} {\rm (#6) #5} ({\em #1})}
\def\hepart[#1]#2{{\rm #2, \em#1}}
\newcommand{\oart}[5]{{\rm #1}, {\em #2 \bf #3} {\rm (#5) #4}}

\newcounter{alphaequation}[equation]
\def\thealphaequation{\theequation\hbox to
0.6em{\hfil\alph{alphaequation}\hfil}}
\def\eqnsystem#1{
\def\@eqnnum{{\rm (\thealphaequation)}}
\def\@@eqncr{\let\@tempa\relax \ifcase\@eqcnt \def\@tempa{& & &} \or
  \def\@tempa{& &}\or \def\@tempa{&}\fi\@tempa
  \if@eqnsw\@eqnnum\refstepcounter{alphaequation}\fi
\global\@eqnswtrue\global\@eqcnt=0\cr}
\refstepcounter{equation} \let\@currentlabel\theequation \def\@tempb{#1}
\ifx\@tempb\empty\else\label{#1}\fi
\refstepcounter{alphaequation}
\let\@currentlabel\thealphaequation
\global\@eqnswtrue\global\@eqcnt=0 \tabskip\@centering\let\\=\@eqncr
$$\halign to \displaywidth\bgroup \@eqnsel\hskip\@centering
$\displaystyle\tabskip\z@{##}$&\global\@eqcnt\@ne
\hskip2\arraycolsep\hfil${##}$\hfil& \global\@eqcnt\tw@\hskip2\arraycolsep
$\displaystyle\tabskip\z@{##}$\hfil
\tabskip\@centering&\llap{##}\tabskip\z@\cr}
\def\endeqnsystem{\@@eqncr\egroup$$\global\@ignoretrue} \makeatother

\newcommand{\SU}{\,{\rm SU}}

\begin{document}
\thispagestyle{empty}
\centerline{IFUP-TH/2015}

\bigskip

\begin{center}
{\LARGE \bf 
\color{verdess}
Totally asymptotically free trinification}\\[1cm]

{\large\bf Giulio Maria Pelaggi, 
Alessandro Strumia {\rm and} Saverio Vignali
}  
\\[7mm]
{\it Dipartimento di Fisica dell'Universit{\`a} di Pisa and INFN, Italy \\[1mm] }
\vspace{1cm}
{\large\bf
Abstract}

\end{center}
\begin{quote}
{\large\noindent
Motivated by new ideas about the Higgs mass naturalness problem, we
present realistic TeV-scale extensions of the Standard Model, into the gauge group $\SU(3)_L\otimes\SU(3)_R \otimes \SU(3)_c$,
such that all gauge, Yukawa and quartic couplings can be extrapolated up to infinite energy.
Three generations of chiral fermions and Higgses are needed, as well as some extra fermion.
}

\end{quote}


\tableofcontents

\setcounter{footnote}{0}


\mysection{Introduction}
The new physics predicted by the Higgs mass naturalness problem 
so far did not show up at the Large Hadron Collider nor in any other experiment.
The  fine-tuning level implied by present bounds is so uncomfortably high 
that  the whole issue is being reconsidered.
One possible alternative approach consists in maintaining the view that nature is natural, but
accepting the possibility that we misunderstood Higgs mass naturalness, attributing physical meaning to quadratically divergent corrections
and insisting on extensions of the SM that cancel them, such as supersymmetry or composite Higgs.

In this work,
we re-interpret naturalness demanding that it is satisfied only by those corrections to the Higgs mass $M_h$
which are physical i.e.\ in principle observable.
Then the SM becomes natural and it is possible to devise natural extensions that include
neutrino masses, Dark Matter~\cite{pap}, gravity~\cite{agravity} and inflation~\cite{ainflation}, as demanded by data.
In general, new physics much above the weak scale is natural provided that it is  coupled weakly enough to the SM:
then the RGE running is dominated by the SM couplings.
In particular, gravity can be a low-energy manifestation of small dimensionless couplings~\cite{agravity,TAF}.


In this context, one would like to have a theory that can hold up to infinite energy~\cite{old,TAF,safe}, 
without any cut-off that could give physical meaning to power divergences.
However, in the SM, 
the hypercharge gauge coupling
$g_Y$ hits a Landau pole around $10^{43}\GeV$.
It is not clear if this implies a correction to the Higgs mass of the same order:
$g_Y$ could reach a non-perturbative interacting fixed point~\cite{Suslov}.
Given that we do not know how to compute this kind of possibility, 
we here assume that all couplings must be asymptotically free.
Then, the requirement that the SM can be extrapolated up to infinite energy without hitting any Landau pole
implies the trivial wrong prediction $g_Y=0$
and non-trivial predictions for $y_t,y_\tau$ and for the Higgs quartic~\cite{TAF}.

In order to have a realistic natural model that satisfies Total Asymptotic Freedom (TAF),  the SM must be extended
around the weak scale into a theory without abelian U(1) factors. 
The specific hypercharges of SM fermions suggest two possibilities~\cite{TAF,safe}:
\beq 
\begin{array}{ll}
\hbox{Pati-Salam:}&G_{224}= \SU(2)_L\otimes\SU(2)_R\otimes\SU(4)_{\rm PS} \\
\hbox{Trinification:} & G_{333}=\SU(3)_L\otimes\SU(3)_R\otimes\SU(3)_c  
\end{array}\ .
\eeq
Asymptotically free gauge couplings are only a first step:  all Yukawa and quartic couplings must also satisfy TAF conditions
described in~\cite{TAF}, where a Pati-Salam TAF model was found.
However the TAF conditions for the quartics
did not allow to realize  Pati-Salam models that avoid quark-lepton unification.
As a consequence, in the TAF Pati-Salam model,
flavor bounds force the masses of
gauge vectors of $\SU(4)_{\rm PS}/\SU(3)_c$ to be heavier than 100 TeV, which is unnaturally above the weak scale.

\smallskip

Trinification~\cite{trini,Pas} does not predict quark-lepton unification and thereby is  safer than Pati-Salam from the point of view of flavour bounds.
Thereby trinification could give rise to simple natural TAF models.
However \cite{TAF} did not find any realistic trinification model that satisfies the TAF conditions.

Since analytic understanding does not offer enough guidance to TAF searches,
in order to perform an extensive brute-force scan, we developed a code 
that, given the gauge group and the field content, 
finds the Yukawa and quartic couplings, computes their one-loop RGE and checks if it admits TAF solutions.


\medskip

In section~\ref{333} we discuss minimal weak-scale trinification, discussing why 3 generations of Higgses are needed.
In section~\ref{search} we find and systematically classify TAF extensions of minimal trinification that only involve extra vector-like fermions.
Results are summarised in the conclusion in section~\ref{concl}.

\begin{table}
$$
\begin{array}{|rccc|ccc|ccc|}\hline
\rowcolor[cmyk]{0,0,0,0.05}
\multicolumn{2}{|c}{\hbox{Field}}&\hbox{spin} &\hbox{generations} 
& \SU(3)_L&\SU(3)_R & \SU(3)_{\rm c}
& \Delta b_L & \Delta b_R & \Delta b_c
\cr \hline
Q_R\,=\!\!&{\small \begin{pmatrix}u^1_R & u_R^2 &  u^3_R  \\ d_R^1 & d_R^2 &  d^3_R  \\    d^{\prime 1}_R & d^{\prime 2}_R & d^{\prime 3}_R\end{pmatrix}}&1/2&3& 1 & 3  & \bar 3 & 0 & 1 &1\cr
Q_L\,=\!\!&{\small \begin{pmatrix}u^1_L & d^1_L &  \bar d^{\prime 1}_R  \\  u_L^2& d_L^2 &  \bar d^{\prime 1}_R   \\  u_L^3& d_L^3 & \bar d^{\prime 3}_R \end{pmatrix}} &1/2&3& \bar 3 &1 &   3 & 1 & 0 &1 \cr
L\,=\!\!&{\small \begin{pmatrix}\bar\nu'_L & e'_L &  e_L \\ \bar e'_L &  \nu'_L & \nu_L  \\  e_R &\nu_R &\nu' \end{pmatrix}} & 1/2& 3&3 & \bar 3  & 1  & 1 &1 & 0\cr
\multicolumn{2}{|c}{H} &0&3  &3 & \bar 3  & 1 & \frac12 & \frac12 & 0 \cr 
 \hline
\end{array}$$
\caption{\em\label{tab:333} Field content of minimal weak-scale trinification.}
\end{table}

\section{Weak-scale trinification}\label{333}

Table~\ref{tab:333} summarises the field content of minimal trinification.\footnote{In most of the literature~\cite{trini}, trinification models include
a permutation symmetry among the 
three $\SU(3)$ factors that forces the trinification scale to be very large.
We do not impose such extra symmetry, partially broken by the scalar field content and
totally broken by the numerical values of the gauge couplings.}

\subsection{Scalars}\label{scalars}
One bi-triplet scalar $H$ in the $(3_L,\bar 3_R)$ representation contains 3 Higgs doublets.
At least two bi-triplets, $H_1$ and $H_2$, are needed in order to break $G_{333}$ to the 
SM gauge group.
Indeed, the most generic vacuum expectation values that give the desired pattern of symmetry breaking 
are
\beq 
\langle H_n\rangle =  \begin{pmatrix} v_{un} & 0&  0 \\   0 &  v_{dn} &  v_{L n}\\  0 & V_{Rn} & V_n\end{pmatrix}.\label{eq:vevs}
\eeq
The vacuum expectation values denoted with a capital $V$ break $G_{333}$ to the SM gauge group 
$G_{123}={\rm U}(1)_{Y} \otimes \SU(2)_L\otimes\SU(3)_c$,
and must be larger than 
the vevs denoted with a lower-case $v$, that break $G_{123}\stackrel{v}{\to}{\rm U}(1)_{\rm em}\otimes \SU(3)_c$.
Notice that $V_{R1}$ and $v_{L1}$ can be set to zero, by redefining the field $H_1$.

\bigskip

The most generic quartic scalar potential is:
\begin{enumerate}
\item $V(H_1)=V_{1111}$ for a single Higgs field $H_1$. It contains two quartic couplings.

\item $V(H_1,H_2) = (V_{1111}+V_{2222})+V_{1122} + (V_{1222}+ V_{1222})$
for two Higgs fields $H_1$ and $H_2$.
It    contains 14 real quartics plus 6 phases.

\item $V(H_1,H_2,H_3) = (V_{1111}+V_{2222}+V_{3333})+
(V_{1122} +V_{2233}+V_{1133})+ (V_{1222}+V_{1333} + V_{2333}+V_{2111}+V_{3111}+V_{3222})+
(V_{1123}+V_{2213}+V_{3312})$
for three Higgs fields $H_1$, $H_2$ and $H_3$.
It contains 54 real quartics plus 36 phases.

\end{enumerate}
We  defined:
\begin{eqnsystem}{sys:V333}
 V_{iiii} &=& \lambda_{ai} \mbox{Tr}(H_i^\dagger H_i)^2 + \lambda_{bi}  \mbox{Tr}(H_i^\dagger H_iH_i^\dagger H_i), \\
 V_{iiij}&=&  \Re   [\lambda_{aiiij}\Tr(H_i^\dagger H_j)\Tr(H_i^\dagger H_i)+  \lambda_{biiij}\Tr(H_i^\dagger H_i H_i^\dagger H_j)] ,  \\
 V_{iijj} &=&  \lambda_{aij} \Tr(H_i^\dagger H_i)\Tr( H_j^\dagger H_j)+   \lambda_{bij} |  \Tr(H_i^\dagger H_j)|^2 
+  \lambda_{cij} \Tr(H_i^\dagger H_i H_j^\dagger H_j)+\\
&&+ \lambda_{dij} \Tr(H_i H_i^\dagger H_j H_j^\dagger)+   \Re[ \lambda_{eij} \Tr(H_i^\dagger H_j)^2 +  \lambda_{fij} \Tr(H_i^\dagger H_j H_i^\dagger H_j)] \, ,\nonumber \\
V_{iijk} & = & \Re \big[\lambda_{aijk}\Tr(H_i H_k^\dagger H_i H_j^\dagger)+
  \lambda_{bijk}\Tr(H_i H_k^\dagger H_j H_i^\dagger)+
  \lambda_{cijk}\Tr(H_i H_i^\dagger H_j H_k^\dagger)+ \\
&& + \lambda_{dijk}\Tr(H_i H_j^\dagger)\Tr(H_i H_k^\dagger)+
  \lambda_{eijk}\Tr(H_j H_i^\dagger)\Tr(H_i H_k^\dagger)+
    \lambda_{fijk}\Tr(H_i H_i^\dagger)\Tr(H_j H_k^\dagger)\big].\nonumber
    \end{eqnsystem}

\subsection{Vectors}
The three  trinification gauge coupling constants ($g_L$, $g_R$, $g_c$) allow to reproduce those of the SM ($g_3$, $g_2$, $g_Y = \sqrt{3/5}\, g_1$) as
\beq g_L = g_2, \qquad  g_R = \frac{2 g_2 g_Y}{\sqrt{3 g_2^2 - g_Y^2}} , \qquad g_c = g_3\, .\eeq

The vev $V_1$ alone
breaks $G_{333}\to \SU(2)_L\otimes\SU(2)_R \otimes{\rm U}(1)_{B-L}\otimes \SU(3)_c$.
At this stage, a left-handed Higgs doublet, a right-handed doubled and one singlet get eaten by the $4+4+1$  vector bosons that acquire masses: a $\SU(2)_L$ doublet $H_L$, a $\SU(2)_R$ doublet $H_R$ and a $Z'$ singlet:
\beq M_{H_L} = {g_L} V_1,\qquad 
 M_{H_R} = {g_R} V_1,\qquad
 M_{Z'} = \sqrt{\frac43 (g_L^2+g_R^2)} V_1.
\label{eq:331br1} \eeq
The massive $Z'$ corresponds to the combination of gauge bosons $g_L A^8_{L\mu} - g_R A^8_{R\mu}$. 
Precision data imply $M_{Z'}> 2-6\TeV$, depending on the $Z'$ charge of the light SM Higgs.

\medskip

Taking into account the $n$ scalars with generic vacuum expectation values $V_n$ and $V_{Rn}$ as in \eq{eq:vevs} and defining 
$V^2\equiv \sum_n (V_n^2+V_{Rn}^2)$
and the dimension-less ratios $\alpha\equiv \sum_n V_{Rn}^2/V^2$ and
$\beta\equiv \sum_n V_n V_{Rn}/V^2$, the gauge bosons form:
\begin{itemize}
\item A left-handed weak doublet with 4 components and mass
$M_{H_L} = g_L V$;

\item The $\SU(2)_R$ vector doublet $H_R$ splits into two charged component with mass
$M_{H_R^\pm}= g_R V$ and into 2 neutral components with mass
\beq  M_{H_R^0}^2 = \frac{g_R^2 V^2}{2} \bigg[ 1 + \sqrt{(1-2\alpha^2)^2+4\beta^2}\bigg]  . \eeq
 
\item The breaking of $\SU(2)_R$ gives rise to right-handed  $W_R^\pm$ vectors with mass
\beq  M_{W_R^\pm}^2 = \frac{g_R^2 V^2}{2} \bigg[ 1 -\sqrt{(1-2\alpha^2)^2+4\beta^2}\bigg]   \eeq
(these are the lightest extra vectors) and to a $Z_R$ vector.

\item The $Z_R$ and the $Z_{B-L}$ vectors mix forming  eigenstates with masses:
\beq M_{Z', Z''}^2 = \frac{2V^2}{3} \bigg[ (g_L^2 + g_R^2)  \pm \sqrt{ (g_L^2 + g_R^2)^2 +3
g_R^2 (4g_L^2+g_R^2) (\alpha^4-\alpha^2+\beta^2 )} \bigg].\eeq
In the limit $V_{Rn}\ll V_{n}$ these reduce to
\beq M_{Z'} = \sqrt{\frac43 (g_L^2+g_R^2)} V,\qquad
M_{B-L} \simeq |\beta| g_RV_R \sqrt{  \frac{g_R^2+4 g_L^2}{g_R^2+g_L^2} } .\eeq

\item  The 12  SM vectors remain massless.
\end{itemize} 
The gauge boson of $B-L$ corresponds to $g_R A^8_{L\mu} + g_L A^8_{R\mu}$ with $g_{B-L} = (\sqrt{3}/2) g_Rg_L/\sqrt{g_R^2+g_L^2}$ and
is subjected 
 to the  bound $M_{B-L}\circa{>}2.6\TeV$ from ATLAS~\cite{TAF}.

\subsection{Fermions}

The SM chiral fermions are contained in a $Q_R\oplus Q_L \oplus L$ multiplet as
described  in table~\ref{tab:333}.
Each generation of $Q_R\oplus Q_L \oplus L$  contains 27  fermions  that decompose under the SM gauge group as the usual 15 SM chiral fermions, plus a vector-like lepton doublet $L'\oplus\bar L'$,
a vector-like right-handed down quark $d'_R\oplus \bar d'_R$, and two neutral singlets, denoted as $\nu_R$ and $\nu'$
in table~\ref{tab:333}.

\medskip

The observed pattern of quark masses is an independent reason why two bi-triplet Higgses $H_1$ and $H_2$ are
needed in order to achieve a realistic model.
In the trinification model with the minimal content of chiral fermions, the SM Yukawa couplings are obtained from the $G_{333}$-invariant interactions
\beq- \Lag_{Y} = y_{Qn}^{ij} ~Q_{Li} Q_{Rj}  H_n + \frac{y^{ij}_{Ln}}{2} L_iL_j H_n +{\rm h.c.}
\label{eqqqY}
\eeq
where summation over $n$ and $i,j=\{1,2,3\}$ is implicit and 
with $y_L$ symmetric under $i\leftrightarrow j$.
The Yukawa couplings satisfy an accidental global U(1) symmetry under which $Q_L$ and $Q_R$ have opposite charges, such that the proton is stable like in the SM.

Expanding in components, and omitting flavor indexes, we find
\beq m_u = v_{un} y_{Qn} ,\qquad
\bordermatrix{& d_R & d'_R\cr
d_L & v_{dn} y_{Qn} & V_{Rn} y_{Qn} \cr
\bar d'_R &  v_{Ln} y_{Qn} & V_n y_{Qn} }
\label{md}
\eeq
for the mass matrices of up-type and down-type  quarks,
\beq\label{me}
\bordermatrix{& e_R & \bar e'_L\cr
e_L &- v_{dn} y_{Ln} & V_{Rn} y_{Ln} \cr
e'_L &  v_{Ln} y_{Ln} &- V_n y_{Ln} }
\eeq
for the charged lepton mass matrix, and
\beq\label{eq:mnu}
\bordermatrix{& \nu_L & \nu_R & \nu'_L & \bar\nu'_L & \nu'  \cr 
\nu_L  &0& - v_{un} y_{Ln} &0 & -V_{Rn} y_{Ln} &0\cr 
\nu_R &&0&0& - v_{Ln} y_{Nn} &0\cr
\nu'_L  &&& 0& V_n y_{Ln}  & v_{un} y_{Ln}\cr
\bar\nu'_L &&&&0 & v_{dn} y_{Ln}\cr
\nu' &&&&&0 \cr
}
\eeq
for the symmetric mass matrix of neutral leptons (`neutrinos') at tree level.
Minimal trinification predicts an {odd} number of neutrinos per generation:
the usual $\nu_L$ with $B-L=-1$;
$\nu_R$ with $B-L=+1$,
and $\nu', \nu'_L,\bar\nu'_L$ with vanishing $B-L$.

%
%
%
%
%
%

%

\subsection{The extra fermions}
The model with $n=2$ Higgs fields is usually considered  as `minimal trinification'; however it has the following problem:
the extra fermions tend to be too light.

Indeed, the extra primed fermions $d'_R$ and $e'_R$ (chiral under $G_{333}$ but not under $G_{\rm SM}$) get masses of order $M' \sim y V$
when $G_{333}$ gets broken by the vev $V$,
and  the chiral SM fermions get masses of order $m\sim yv$ such that $M'/m \sim V/v$.
More precisely, from the mass matrices of \eq{md} and \eq{me}, one finds the masses of the heavy extra fermions
\beq
M_{d'_R} \approx \sqrt{(V_n y_{Qn})^2+(V_{Rn} y_{Qn})^2} ,\qquad
M_{e'_R} \approx \sqrt{(V_n y_{Ln})^2+(V_{Rn} y_{Ln})^2} \eeq
and of the light SM fermions:
\beq
 \label{Md'R1g}\qquad
m_d \approx \frac{(v_{dn} y_{Qn} )(V_n y_{Qn}) - (v_{Ln} y_{Qn})(V_{Rn} y_{Qn})}{M_{d'_R}},\quad
m_e \approx \frac{(v_{dn} y_{Ln} )(V_n y_{Ln}) - (v_{Ln} y_{Ln})(V_{Rn} y_{Ln})}{M_{e'_R}}.
\eeq
Thereby, the masses of first generation quarks and leptons are naturally reproduced for Yukawa couplings of order $y\circa{<} 10^{-5}$, like in the SM.
Notice that the two Yukawa couplings $y_{Q1}$ and $y_{Q2}$ allow to reproduce the two masses of the SM up and down quarks.

The problem is that TeV-scale minimal trinification with $V \sim \hbox{few TeV}$ naturally implies extra fermions with masses $M'\circa{<} 0.1\GeV$, 
in sharp contradiction with data.  In particular, $M_{d'_R}$ is a few thousand times below its LHC bound $M_{d'_R} \circa{>} 700\GeV$~\cite{d'LHC}.
Second and third generation quarks are heavier, giving rise to a qualitatively similar but quantitatively smaller problems with $s'_R$, $b'_R$.

Comparing the number of experimental constraints to the number of free parameters (and including the vevs $v$ among them)
shows the existence of experimentally allowed but fine-tuned choices of parameters, such that $M'$ is heavy enough.
However, the fine-tuning needed to avoid all problems in all generations is at the $10^{8}$ level: we do not want to purse this road.

\bigskip

\bigskip

A simple way out is considering weak-scale trinification  models with 3 Higgses $H_{1,2,3}$.
The three Yukawa couplings $y_{Q1},y_{Q2},y_{Q3}$  then allow to naturally adjust the three masses $m_u$, $m_d$ and $M_{d'_R}$
to values compatible with experiments.

For example, a natural configuration is obtained assuming that $H_1$ breaks $G_{333}$ but preserves $G_{\rm SM}$
(i.e.\ $V_1\neq0$ and $v_{d1}=v_{u1}=v_{L1}=0$).  Then, the Yukawa couplings $y_{Q1}$ and $y_{L1}$ allow to give large enough 
masses $M_{d'_R} =V_1 y_{Q1}\circa{>}700\GeV$ and $M_{e'_R}=V_1 y_{L1}\circa{>}200\GeV$ to the extra primed fermions,
without also giving too large masses to the SM fermions.
The other Higgses $H_2$ and $H_3$ can have the small Yukawa couplings needed
to reproduce the light SM fermion masses,
\beq m_e \sim \sum_{n=2}^3  v_{dn} y_{Ln},\qquad m_u\sim \sum_{n=2}^3 v_{un} y_{Qn}, \qquad
m_d\sim \sum_{n=2}^3 v_{dn} y_{Qn}.\eeq

\medskip

\subsection{Neutrinos}\label{nu}
So far we ignored neutrinos, which deserve a special discussion.
Finite naturalness demands that neutrino masses be generated at relatively low energy~\cite{pap}, as in trinification models
where U$(1)_{B-L}$ is gauged and gets spontaneously broken at $V_R\sim \hbox{few TeV}$.
The neutrino mass matrix of \eq{eq:mnu} contains, in its 12 component, the Dirac entry $ \sum_{n=2}^3 v_{un} y_{Ln}$, which can naturally be
much smaller than $m_u$ (which involves the same $v_{un}$)
and than $m_e$ (which involves the same $y_{Ln}$) if one assumes that $y_{L3}$ is small and that $v_{u2}=0$.
Notice however that the one-loop RGE running of $y_{L3}$ implies that it cannot be arbitrarily small:
\beq
(4\pi)^2\frac{d y_{L_3}}{d\ln\mu} = 6 y_{L_3}^3 + y_{L_3}(-8 g_c^2-4g_L^2-4g_R^2+6 y_{L_1}^2+6y_{L_2}^2) + 3 y_{Q_3}(y_{Q_1}y_{L_1}+y_{Q_2}y_{L_2}).
\eeq
The tree-level neutrino mass matrix of \eq{eq:mnu}, in the $G_{\rm SM}$-preserving limit $v=0$,
gives rise to 3 massless eigenstates (1 active, 2 sterile) per generation.
So, at tree level, extra sterile neutrinos  remain light despite not being chiral under the SM.
This is no longer true at one loop level:   the extra sterile neutrinos acquire Majorana masses of order $M \sim V y_L^3/(4\pi)^2$,
leaving light active neutrinos with masses $m_\nu \sim (v_{un} y_{Ln})^2/M$.
Ref.~\cite{Pas} presented regions of parameters space where neutrinos are naturally light.

Majorana mass terms can also be obtained at tree level adding extra fields with mass $M\sim V$, 
such that the observed neutrino masses are
reproduced for $y_{Ln}\sim 10^{-6}$
One of the two extra sterile neutrinos per generation becomes massive
adding one extra fermion singlet $N_i$ per generation, with Majorana mass $M_N$:
\beq \Lag_{\rm extra} = \bar N_i i \slashed{\partial} N_i -y_{Nn}^{ij} N_i L_j H^*_n - \frac{M_N^{ij}}{2} N_i N_j  . \eeq
Indeed, integrating out $N$ generates the operator $\Tr(LH^\dagger)^2$, which gives a Majorana mass to one combination of light sterile neutrinos.
Both extra sterile neutrinos become massive adding Majorana fermions $8_L$ in the adjoint of $\SU(3)_L$ and/or $8_R$ in the adjoint of $\SU(3)_R$:
integrating them out generates $\Tr(L T^a H^\dagger)^2$ operators.
The addition of these fields is not only compatible with Total Asymptotic Freedom but (in some models) also necessary, as discussed below.


\section{Totally Asymptotically Free trinification}\label{search}
A first search for TAF trinification models was conducted in~\cite{TAF}, finding only TAF models with a single Higgs $H_1$.
As discussed in the previous section $n=2$ Higgses $H_{1}$ and $H_2$ are needed for a fine-tuned trinification weak-scale model,
and $n=3$ Higgses $H_{1},H_2,H_3$ are needed for a natural weak-scale trinification model.

The purpose of this section is finding trinification TAF models with $n>1$ Higgses, despite that,
as discussed in section~\ref{scalars},
the number of quartic couplings that must satisfy TAF conditions grows from 2 ($n=1$) to 20 ($n=2$) to 90 ($n=3$).

First, we consider minimal trinification.  It has asymptotically free gauge one-loop $\beta$-functions 
\beq \label{eq:betag}
 \frac{dg_i}{d\ln\mu} = b_i \frac{  g_i^3}{(4\pi)^2},\qquad
b_L = b_R = -5+\frac{n}{3},\qquad
b_c =  -5
.\eeq
However, we find that the quartics of minimal trinification  do not satisfy the TAF conditions
when $n>1$ Higgses are present.

We then perform a systematic analysis of all vector-like fermion multiplets that can be added to 
minimal trinification keeping all gauge couplings $g_L,g_R,g_c$
asymptotically free.
We do not explore the possibility of adding extra scalars beyond the $n$ Higgses.
We find  3035 possible combinations  of extra fermions for $n=2$ (and slightly less for $n=3$), to which one can add
any number of singlets under $G_{333}$.

\medskip

\begin{table}[t]
$$
\begin{array}{c|cc|ccc|cc}
\rowcolor[cmyk]{0,0,0,0.05}
&\hbox{name} & \hbox{representation} & \multicolumn{3}{c}{\Delta b_i} & \multicolumn{2}{|c}{\rm Yukawas}\\ \hline
\parbox[t]{3mm}{\multirow{3}{*}{\rotatebox[origin=c]{90}{unstable\hspace{8mm}}}} 
& 1  & (1,1,1) & 0&0&0& 1 LH^* & -\\
& 8_L & (8,1,1) &2&0&0  & 8_L L H^* & -\\
& 8_R & (1,8,1) & 0 &2&0   & 8_R L H^* & - \\
& L'  \oplus \bar L'& (3,\bar 3,1)\oplus(\bar 3, 3,1) &2&2&0   &  L'LH &  L'L'H + \bar L' \bar L' H^* \\
& Q'_L  \oplus \bar Q'_L& (\bar 3,1,3)\oplus(3,1,\bar 3) & 2&0&2 & Q'_L Q_R H& -\\
& Q'_R\oplus\bar Q'_R & (1,3, \bar 3)\oplus (1,\bar 3,3) & 0&2&2 & Q'_R Q_L H& - \\ \hline 
\parbox[t]{3mm}{\multirow{3}{*}{\rotatebox[origin=c]{90}{stable\hspace{25mm}}}} 
& 3_L \oplus \bar 3_L& (3,1,1)\oplus(\bar 3,1,1) & \frac23 & 0 &0& -& - \\
& 3_R\oplus \bar 3_R & (1,3,1)\oplus (1,\bar 3,1) & 0 & \frac23 &0& -& - \\
& 3_c\oplus \bar 3_c& (1,1,3)\oplus(1,1,\bar 3)  & 0 &0&  \frac23 & -& - \\
& 8_c & (1,1,8) & 0&0&2& -& -\\
& 6_L \oplus \bar 6_L& (6,1,1)\oplus(\bar 6,1,1)& \frac{10}{3} &0&0& -& - \\
& 6_R \oplus \bar 6_R& (1,6,1)\oplus(1,\bar 6,1) & 0 & \frac{10}{3} &0& -& - \\
& 6_c\oplus \bar 6_c& (1,1,6)\oplus(1,1,\bar 6) & 0 &0& \frac{10}{3}& -& -  \\
& \tilde L \oplus \bar {\tilde{L}}& (3,3,1) \oplus (\bar 3,\bar3,1) & 2&2&0& -& -\\
& \tilde Q_L\oplus\bar{\tilde Q}_L & (3,1,3) \oplus (\bar 3,1,\bar3) & 2&0&2& -& -\\
& \tilde Q_R \oplus\bar{\tilde{Q}}_R& (1,3,3) \oplus (1,\bar 3,\bar3) & 0&2&2& -& -\\
\end{array}
$$
\caption{\em Fermionic multiplets that can be added to minimal trinification while keeping gauge couplings asymptotically free.\label{list1}}
\end{table}

\begin{table}[t]
$$
\begin{array}{c|cccccc}
 \diaghead(-1,1){\theadfont aaaaa}{$b_L$}{$b_R$}&-4   & -10/3 & -8/3 &  -2  & -4/3 & -2/3\\
\hline
-4   &-3   & -7/3  & -7/3 & -7/3 & -5   & -5  \\
-10/3&-7/3 & -7/3  & -7/3 & -5/3 & -5   & -5 \\
-8/3 &-7/3 & -7/3  & -7/3 & -5/3 & -5   & -5 \\
-2   &-7/3 & -5/3  & -5/3 & -5/3 & -5   & -5 \\
-4/3 & -5  & -5    & -5   & -5   & -5  & -5 \\
-2/3 & -5  & -5   & -5  & -5  & -5  & -5 
\end{array}
$$
$$
\begin{array}{c|cccccc}
\diaghead(-1,1){\theadfont aaaaa}{$b_L$}{$b_R$}
       &-7/2   & -17/6 &-13/6 & -3/2 & -5/6 & -1/6\\
\hline
-7/2   &-5/3   & -5/3  & -5/3 & -1 & -5   & -5  \\
-17/6  &-5/3   & -5/3  & -1   & -1 & -5   & -5 \\
-13/6  &-5/3   & -1    & -1   & -1 & -5   & -5 \\
-3/2   & -1    & -1    & -1   & -5 & -5   & -5 \\
-5/6   & -5    & -5    & -5   & -5 & -5   & -5 \\
-1/6   & -5    & -5    & -5   & -5 & -5   & -5
\end{array}
$$
\caption{\em Most negative value of the $\SU(3)_c$ one-loop $\beta$-function coefficient 
$b_c$ as function of $b_L$ and $b_R$ (analogous coefficients for $\SU(3)_L$ and $\SU(3)_R$) 
such that minimal trinification plus extra fermions without Yukawa couplings
satisfies Total Asymptotical Freedom.
The upper (lower) table applies to $n=2$ ($n=3$) Higgses, considering in each case the discrete grid of allowed values of $b_L$ and $b_R$.
\label{tabb}}
\end{table}

\subsection{TAF models with extra stable fermions}

In order to analyse in a systematic way this large number of possibilities, we
list  in table~\ref{list1} the allowed extra fermions: the models are obtained adding combinations of them.
There are 16 possible kinds extra fermions, that split into two categories: stable and unstable.
The first 6 fermions are  unstable
because they can have Yukawa couplings with the SM chiral fermions
that induce their decays.
The latter 10 fermions cannot have such Yukawa couplings, because of group theory and renormalizability.
In some combinations, such `stable' fermions can have Yukawa couplings among themselves, but 
 the Lagrangian still accidentally satisfies a Z$_2$ symmetry under which their sign is reversed.
Thereby, at least the lightest component of the fermions belonging to the `stable' category
is stable.  

\medskip

All components of the extra stable fermions are either charged or colored.
For example, fermions in the $3_L$ and $3_R$ representations have the same fractional charge as the quarks in $Q_L$ and $Q_R$, but without color.
This means that they are not good Dark Matter candidates, and that the bounds on their cosmological abundance are so strong
that the temperature of the universe  must  have always been well below their masses,
that must be lighter than a few TeV for naturalness reasons~\cite{pap}.
While this is not excluded, we will not purse this possibility, apart from
showing which combinations of the stable 10 multiplets provide TAF solutions.

\smallskip

Given that the extra fermions do not have Yukawa couplings with the SM fermions
(and neglecting possible Yukawa couplings among pairs of extra fermions),
TAF solutions can only appear as long as the addition of the extra fermions makes the gauge $\beta$ functions closer enough to 0.
Table~\ref{tabb} shows, as function of $b_L$ and of $b_R$,
the lowest value of $b_c$ such that TAF solutions are found.\footnote{Such TAF solutions were not found in~\cite{TAF},
because there all Yukawa couplings of the 1st and 2nd generations were neglected,
a simplifying assumption relaxed in the present study.}
Table~\ref{tabb} dictates which combinations of extra stable fermions lead to TAF models:
for example, its upper-left entry ($b_c = -3$ for the minimal values of $b_L=b_R=-4$ and for $n=2$) means that TAF solutions are obtained increasing $b_c^{\rm minimal} = -5$ by $\Delta b_c = 2$ (as realised adding, for example, a gluino-like fermion in the adjoint of $\SU(3)_c$).
The results for $n=3$ Higgses are similar to those for $n=2$.

\medskip

\begin{table}[t]
$$\hspace{-0.5ex}
\begin{array}{c|ccc|ccc}
\rowcolor[cmyk]{0,0,0,0.05}
\hbox{Extra fermions} & b_L  & b_R & b_c& \multicolumn{2}{|c}{\rm Extra~Yukawas} & \hbox{TAF} \\ \hline
8_L  &-3+\frac{n}{3} &-5+\frac{n}{3}&-5&   8_L L H^* &&\hbox{yes$^*$}\\
8_R  &-5+\frac{n}{3}&-3+\frac{n}{3}&-5& 8_R L H^* &&\hbox{yes$^*$}\\
8_L \oplus 8_R  &-3+\frac{n}{3}&-3+\frac{n}{3}&-5&   8_L L H^*+8_R L H^* &&\hbox{yes$^*$} \\
L'  \oplus \bar L'   &-3+\frac{n}{3}&-3+\frac{n}{3}&-5&    L'LH &  L'L'H + \bar L' \bar L' H^*&\hbox{yes}^* \\
Q'_L  \oplus \bar Q'_L &-3+\frac{n}{3}&-5+\frac{n}{3}&-3&   Q'_L Q_R H &&\hbox{no} \\
Q'_R\oplus\bar Q'_R   &-5+\frac{n}{3}&-3+\frac{n}{3}&-3& Q'_R Q_L H&&\hbox{no} \\ 
Q'_L  \oplus \bar Q'_L \oplus 8_R &-3+\frac{n}{3}&-3+\frac{n}{3}&-3& Q'_L Q_R H + 8_R L H^* & &\hbox{yes}\\
Q'_R  \oplus \bar Q'_R \oplus 8_L &-3+\frac{n}{3}&-3+\frac{n}{3}&-3& Q'_R Q_L H + 8_L L H^* &&\hbox{yes}\\
Q'_L \oplus Q'_R \oplus \bar Q'_L \oplus\bar Q'_R &-3+\frac{n}{3}&-3+\frac{n}{3}&-1& Q'_L Q_R H + Q'_R Q_L H & Q'_LQ'_RH+\bar Q'_L \bar Q'_R H^* & \hbox{yes} \\
\end{array}
$$
\caption{\em Candidate TAF trinification models with extra unstable fermions.  Any number of singlets can be added to any model.
``Yes$^*$'' means that TAF solutions need  non-vanishing asymptotic fixed-points for the
Yukawa couplings of lighter generations.
This table holds for both $n=2$ and $n=3$ Higgses. 
\label{list2}}
\end{table}

\subsection{TAF models with extra unstable  fermions}
We focus on models containing only combinations of the 6 unstable extra  fermionic multiplets listed in table~\ref{list1}.
By grouping them in all possible ways, we find only 9 combinations that keep
$b_L,b_R,b_c <0$: such 9 candidate TAF models are listed in table~\ref{list2}.

We find that most models satisfy all TAF conditions, as described in the last row of table~\ref{list2}.
In a Mathematica file attached to this paper we show the RGE and some TAF solutions for the first model.
The fixed point solutions form a complicated continuum, as dictated by the 
U(3) global flavor rotations acting on fermions, by the U($n$) `slavor' (scalar flavor) global rotations acting on scalars,
and by their subgroups.

\bigskip


A potential problem\footnote{
We thank Marco Nardecchia and Luca di Luzio for having raised this issue.}
 is that the two-loop gauge $\beta$ functions
could become dominant at large gauge coupling and have opposite sign
to the accidentally small one-loop gauge $\beta$ functions, 
making impossible to run reaching the observed value of the strong gauge coupling $g_c$,
which is the biggest gauge coupling.
The two-loop  RGE for the strong gauge coupling in presence of $n_F$ fermionic fundamentals of $\SU(3)_c$ is
\beq \frac{dg_c}{d\ln\mu} = (-11+\frac{n_F}{3})  \frac{g_c^3}{ (4\pi)^2}+(-102 + \frac{19}{3} n_F)  \frac{g_c^5}{ (4\pi)^2}+\cdots .\eeq
We found TAF models for $18\le n_F \le 30$: in all this range
the two loop term is  subdominant enough that the physical value, $g_c(\bar\mu=3\TeV )=1$, can be reached.

\medskip

Another potential problem is bounds from flavor-violating experiments.
Such bounds are satisfied for new particles in the few TeV mass range, provided that 
the various mixing matrices have off-diagonal entries as small as the CKM mixing matrix~\cite{TAF}.
This level of smallness is respected by quantum corrections such as RGE evolution, and is thereby natural.
Lepton-flavor violating processes can be similarly confined to the small Yukawa couplings needed to generate neutrino masses.

The collider phenomenology of TAF models involving only extra unstable leptons in the $8_L$ and/or $8_R$ representations
is very similar to minimal trinification.   The addition of these extra leptons makes easier to reproduce the observed
neutrino masses, as described in section~\ref{nu}.
The addition of one extra $Q_L\oplus\bar Q_L$ and/or $Q_R\oplus\bar Q_R$ gives rise to larger quark mass matrices.
One would expect that their addition allows to reduce the fine-tuning needed to have $M_{d'\_R}\gg m_d$ present in models with only $n=2$ Higgses.
However the explicit form of the down-quark mass matrix, written in appendix~\ref{QLQR}, shows that this is not the case.

The number of Yukawa and quartic couplings univocally predicted by the TAF conditions because their flows are IR-attractive
can be easily computed for each fixed point following the procedure in~\cite{TAF} and typically is half of all the couplings.
However, in order to extract the resulting physical predictions, a numerical
study of RGE running down to the weak scale and of the minimisation of the potential is needed.
This goes beyond the scope of the present paper.

 \bigskip

Finally, we comment about naturalness.
The squared mass of the SM Higgs doublet receives quantum corrections proportional to the squared masses of the heavy vectors.
The order one factors depend on how the SM Higgs doublet lies inside the trinification multiplets $H_i$.
For example, assuming that it lies in a doublet under $\SU(2)_R$, one has the contribution due to $\SU(2)_R$ heavy vectors 
\beq \delta M_h^2 = - \frac{3g_R^2 M_{W_R}^2}{(4\pi)^2} \bigg[3\ln(\frac{M_{W_R}^2}{\bar\mu^2})+c\bigg] \eeq
where $c$ is an order-one scheme dependent factor.
Approximating the factor in square brackets as unity, naturalness demands $M_{W_R}\circa{<}2\TeV\times \sqrt{\Delta}$ where $\Delta$ is the usual fine-tuning factor.  
Furthermore, TAF conditions typically give quartic couplings of order $g^2$, such that all Higgses typically have comparable masses
and (barring special structures) a tree-level fine-tuning of order $\Delta \sim (V/v)^2$ is needed to have $v\ll V$.

\section{Conclusions}\label{concl}
Motivated by new ideas about the Higgs mass hierarchy problem,
we searched for realistic weak-scale  extensions of the Standard Model that can be extrapolated up to infinite energy i.e.\ 
models
where all gauge and all Yukawa and quartic couplings can run up to infinite energy without hitting any Landau pole,
realising Total Asymptotic Freedom (TAF).

\medskip

A promising candidate is trinification models, based on the gauge group $\SU(3)_L\otimes\SU(3)_R\otimes\SU(3)_c$
broken to the SM at the scale $V$.
In section~\ref{333} we discussed trinification models, confirming that at least 
$n=2$ generations of Higgses are needed to reproduce all observed lepton and quark masses $m$.
However, trinification predicts extra fermions with mass $M' \sim m V/v$
(where $v=174\GeV$ is SM Higgs vev), which are too light for $V\approx \hbox{few TeV}$,
unless fine-tunings are introduced to make them heavier.
We found that these fine-tunings are avoided in the presence of $n=3$ generations of Higgses.

\medskip

In section~\ref{search} we performed a systematic search of TAF trinification models
obtained adding extra vector-like fermions to Minimal Trinification models with $n=2$ or $n=3$ Higgses.  
Previous searches~\cite{TAF} only found TAF solutions for unrealistic trinification models with $n=1$:
the number of quartic couplings that must satisfy TAF conditions is 2 for $n=1$, 20 for $n=2$, $90$ for $n=3$.

We succeeded in finding trinification TAF models for both $n=2$ and $n=3$.
We found many models that predict extra charged or colored stable particles, 
as well as various models, listed in table~\ref{list2},  that do not predict any exotic nor stable extra particles.
About half of the Yukawa and quartic couplings are univocally predicted, because their corresponding fixed flows are IR-attractive.

\medskip


These models are interesting also from other points of view:
trinification explains the observed quantised hypercharges;
the extra $W_R$ vectors can fit the di-boson anomaly present in LHC run I data~\cite{anomalies}
at $M_{W_R}\approx 1.9\TeV$; trinification predicts $g_R/g_L\approx 0.67$, as favoured by the anomaly, and $M_{Z'}\approx 1.7 M_{W_R}$.


\small

\appendix

\section{The trinification TAF model with extra $Q_L$ and  $Q_R$}\label{QLQR}
Trinification models that satisfy all TAF conditions are obtained adding to the minimal trinification model
(containing the chiral fermions $Q_L^i, Q_R^i$ and $L^i$)
one extra vector-like family of quarks
 $Q_L\oplus \bar Q_L$ and/or $Q_R\oplus \bar Q_R$.
The most generic fermion Lagrangian now contains fermion mass terms $M_L$ and $M_R$ and extra Yukawa couplings:
\beq- \Lag_{Y} = M_L^{i'} Q_{Li'}\bar Q_L + M_R^{j'} Q_{Rj'}\bar Q_R+
y_{Q}^{ni'j'} ~Q_{Li'} Q_{Rj'}  H_n
+y_{\bar Q}^n ~\bar Q_L \bar Q_R  H^*_n 
+ \frac{y_{L}^{nij}}{2} L_iL_j H^{*}_n +{\rm h.c.}
\label{eqqqY}
\eeq
where now $i',j'=\{1,2,3,4\}$.
The scalar quartic potential remains as in section~\ref{scalars}.
%
The quark Yukawa matrix  is
\beq \bordermatrix{
& Q_{Rj}  &Q_{R4} & \bar Q_L \cr
Q_{Li} & y_{Q}^{nij} H_n & y_Q^{ni4} H_n & M_L^i\cr
Q_{L4} & y_Q^{n4j} H_n & y_{Q}^{n44} H_n & M_L^4\cr
\bar Q_{R}  & M_{R}^j & M_R^4 & y^n_{\bar Q} H^*_n
}.
\eeq
One can always choose a basis where $M_L^i = M_R^j=0$, such that $M_L^4=M_L$ and $M_R^4 = M_R$.
Inserting the vacuum expectation values, 
and writing 
$Q_{Li} = (u_{Li},d_{Li},\bar d'_{Ri})$,
$Q_{L4}=(U_L,D_L,\bar D'_R)$,
$\bar Q_{L4} = (\bar U_L,\bar D_L, D'_L)$,
$Q_{Rj}=(u_{Rj},d_{Rj},d'_{Rj})$, 
$Q_{R4}=(U_R, D_R, D'_R)$,
$\bar Q_{R4} = (\bar U_R, \bar D_R, \bar D'_L¤)$,
the mass matrices are
\beq
\bordermatrix{
  & u_{Rj} & U_R & \bar{U}_L \cr
 u_{Li} & v_{un} y_Q^{nij} & v_{un} y_Q^{ni4}& 0 \cr
 U_L & v_{un} y_Q^{n4j} & v_{un} y_Q^{n44} & M_L \cr
 \bar{U}_R & 0 & M_R & v_{un} y_{\bar{Q}}^n }
\eeq
for the up-quarks, and
\beq
\bordermatrix{
  & d_R^j & d_R^{j'} & D_R' & D_R & \bar D_L' & \bar{D}_L \cr
 d_L^i & v_{dn} y_Q^{nij} & v_L y_Q^{2ij} & v_L y_Q^{2i4} &  v_{dn} y_Q^{ni4}& 0 & 0 \cr
 \bar{d}_R^{i '} & V_R y_Q^{2ij} &  V_n y_Q^{nij} & V_n y_Q^{ni4} & V_R y_Q^{2i4} & 0 & 0 \cr
 \bar{D}_R' & V_R y_Q^{24j} & V_n y_Q^{n4j} & V_n y_Q^{n44} & V_R y_Q^{244} & M_L & 0 \cr
 D_L & v_{dn} y_Q^{n4j} & v_L y_Q^{24j} & v_L y_Q^{24j} & v_{dn} y_Q^{n44}& 0 & M_L
   \cr
 {D}_L^{'} & 0 & 0 & M_R & 0 & v_{un} y_{\bar{Q}}^n & 0 \cr
 \bar{D}_R & 0 & 0 & 0 & M_R & 0 & v_{un} y_{\bar{Q}}^n }
\eeq
for the down quarks.
A non-trivial feature of this tree-level mass matrix is that the
Yukawa couplings of the extra vector-like fermions do not provide masses of order $V$
for the extra $d'_R$ quarks, such that $n=3$ Higgses remain necessary in order to obtain a model
that naturally satisfies experimental constraints on their masses while reproducing observed quark masses.

\end{document}